\documentclass[11pt]{article}
\usepackage[english]{babel}
\usepackage[T1]{fontenc}
\usepackage{amsmath,amsfonts}
\usepackage{graphicx}
\usepackage[hidelinks]{hyperref}
\usepackage{braket}
\usepackage{color}
\usepackage{soul}
\usepackage{subcaption} 
\usepackage[utf8]{inputenc}
\usepackage{algorithm}
\usepackage{algpseudocode}
\floatname{algorithm}{Procedure}

\usepackage[export]{adjustbox}

\begin{document}
	
	\begin{center}
		
	{\Large{On the existence of nematic-superconducting states in the Ginzburg-Landau regime}}
		
		\bigskip
		
	Mariano De Leo$^1$, Juan Pablo Borgna$^2$, Diego Garc\'ia Ovalle$^3$

        \medskip
        
        \small{$^1$INMABB (CONICET) - Dept. of Math,  UNS, Arg.}\\
        \small{$^2$ICIFI (CONICET) - Applied Math Center, UNSAM - Arg.}\\
        \small{$^3$Aix-Marseille Université, CNRS, CINaM, Marseille, France.}

	\end{center}
	
	\begin{abstract}
		In this article, we investigate the existence of nematic-superconducting states in the Ginzburg-Landau regime, both analytically and numerically. From the configurations considered, a slab and a cylinder with a circular cross-section, we demonstrate the existence of geometrical thresholds for the obtention of non-zero nematic order parameters. Within the frame of this constraint, the numerical calculations on the slab reveal that the competition or collaboration between nematicity and superconductivity is a complex energy minimization problem, requiring the accommodation of the Ginzburg-Landau parameters of the decoupled individual systems, the sign of the bi-quadratic potential energy relating both order parameters and the magnitude of the applied magnetic field. Specifically, the numerical results show the existence of a parameter regime for which it is possible to find mixed nematic-superconducting states. These regimes depend strongly on both the applied magnetic field and the potential coupling parameter. Since the proposed model corresponds to the weak coupling regime, and since it is a condition on these parameters, 
		we design a test to decide whether this condition is fulfilled.  
	\end{abstract}
	
	\section{Introduction}
	The connection between symmetry-related phenomena and superconductivity can produce enormous benefits in nano-electronics. Attention has been devoted on answering whether a combination of superconductivity and spontaneous rotational symmetry breaking can be induced in realistic materials, and ultimately, on its fundamental origin and how it behaves under external impulses. In this context, the so-called nematic superconductivity (NSC) has been previously identified in a wide variety of quantum materials \cite{Huang2018,Andersen2018, Liu2019, Hardy2019, Hardy2019dos, Islam2021,Yang2022,Chichinadze2020,Kushnirenko2020,Cho2022,Siddiquee2022}, including doped topological superconductors \cite{Shen2017,Yonezawa2017,Tao2018,Willa2018,Kuntsevich2019,Sun2019,How2019,Kostylev2020}. Indeed, NSC emerges as a result of the presence of hexagonal warping in topological insulators \cite{Akzyanov2020}, and in doped topological systems, the Lifshitz transition and the shape of the Fermi surface affect NSC, resulting in an intermediate impact against disorder \cite{Akzyanov2021}. This scenario has been investigated in the specific instance of $\textrm{Bi}_2\textrm{Se}_3$ in the presence of magnetization \cite{Khokhlov2021}. The topological properties of 2D NSCs can be characterized through thermal transport, which is a result of the variation of the superconducting gap when nematic fluctuations are present \cite{Choudhoury2022}. In this frame, it has been hypothesized that electromagnetic \cite{Gali2022} and odd-parity \cite{Wu2017} fluctuations play a role in the occurrence of NSC in a general class of unconventional superconductors.\bigskip 
	
	From a theoretical standpoint, the novel studies on $\textrm{Cu}_x\textrm{Bi}_2\textrm{Se}_3$ \cite{Fu2010, Fu2016, Hao2017}, for instance, inspired further investigations on the role of the rotational symmetries of the crystal, the possible identification of NSC through the upper critical field and the interplay between the order parameters in multi-component superconductors. The description of the aforementioned functions was introduced in systems such as \textrm{FeSe} \cite{Kang2018}, which actually displays time-reversal breaking NSC \cite{Kang2018dos}. Nonetheless, recent experiments reveal that this iron-based superconductor exhibits a momentum-dependent nematic order parameter \cite{Pfau2019}.\bigskip 
	
	The electromagnetic response of these materials has been correlated to the influence of chiral Higgs modes \cite{Uematsu2019}. In doped topological systems such as $\textrm{M}_x\textrm{Bi}_2\textrm{Se}_3$ (\textrm{M=Cu, Sr, Nb}) these excitations are accompanied by the vibrations of the nematic order parameter, but their origin is related to the presence of chiral Cooper pairs \cite{Uematsu2019}. Under the effect of visible light, the appearance of the NSC state has also been measured in the iron-based superconductor $\textrm{Ba}_{1-x}\textrm{K}_x\textrm{Fe}_2\textrm{As}_2$\cite{Grasset2022}. Besides, it has been conjectured that NSC can be stabilized by density-wave fluctuations \cite{Fu2019}, addressing its potential role in twisted bilayer Graphene. From an experimental point of view, nematic fluctuations have been verified by using Raman spectroscopy in the cuprate superconductor $\textrm{Bi}_2\textrm{Sr}_2\textrm{CaCu}_2\textrm{O}_{8+\delta}$ \cite{Auvray2019}, where the Pair-Density Wave state (PDW) has also been measured \cite{Hamidian2016}.\bigskip 
	
	In view of the remarkable interest in these materials, for instance, on the competition between the nematic and the superconducting order parameters \cite{Moon2012,Wang2013,Kim2014,Kaczmarczyk2016}, in this manuscript we apply the Ginzburg-Landau theory for the Nematic Superconductor state (GLNSC) \cite{Barci2016} to study the existence of nematic-superconducting states (denoted here as \textit{mixed solutions}), both analytically and numerically. The model is analyzed without taking into account any vortical effects, putting the emphasis on the spatial variation of the superconducting and nematic densities $f(\Vec{x})$ and $g(\Vec{x})$, respectively. We focus on the interaction between the two order parameters in the case where the potential has a weak quartic coupling term, referred in \cite{Barci2016} as the \textit{weak coupling condition}, including in our study both the formal derivation of the model and the numerical treatment of its solutions. Since our computations show that the region in the $(v,\beta)$ plane containing \textit{mixed solution} is rather small, the first challenge is to find the correct regime of parameters. The second one is to test whether the chosen parameters $(v,\beta)$ lead to a numerical solution that is consistent with the weak coupling condition. Throughout this analysis, we compare our results with previous work \cite{Fernandes2021,Severino2022,Bannikov2021}, which considers a model for superconducting materials with a real phase parameter that also neglects vortical effects. \bigskip 
	
	This paper is organized as follows: In Section \ref{SecIIA} we present the general Helmholtz Free Energy and the corresponding Ginzburg-Landau equations. Then we pay attention on two special instances: in Section \ref{SecIIB} we consider a slab in Section \ref{SecIIC} we take an infinite cylinder with a circular cross-section. Both geometries are subjected to an external magnetic field. In Section \ref{III} we deduce the existence of a Fréedericksz threshold of geometric nature for each case. Furthermore, we obtain an estimate involving the coherence length and the typical dimensions of the geometries considered. In Section \ref{IV} we report the numerical results for the case of the slab, which are mainly based on the Shooting Method, focusing our attention to the values of the parameters for which we obtain a mixed response. Then, we compare our outcomes to what is expected from the physical point of view. Finally, in Section \ref{V} we give the main conclusions of our study.   
	
	\section{Derivation of the Model}\label{SecII}
	In this Section, we obtain a system of equations that models the response of a superconducting nematic medium under the presence of an external uniform magnetic field. In this setting, we assume that the model parameters are independent of temperature, keeping the attention only on their spatial profile and the magnetic field. Indeed, although the Ginzburg-Landau theory has been applied to explain successfully numerous effects where NSCs are involved \cite{Barci2016,Fernandes2021,Severino2022}, it has been demonstrated that it is not appropriate when the anisotropy is temperature dependent \cite{Bannikov2021}.
	
	\subsection{Helmholtz Free Energy and derivation of the general model}\label{SecIIA}
	Let us consider the Helmholtz Free Energy per unit length \cite{Barci2016}:
	\begin{eqnarray}
		F_L&=&\iint_{\Omega} V \left(\psi(x),Q(x)\right)d\vec{x}\nonumber\\
		&+&\iint_{\Omega}\left(\alpha_s|\vec{D}\psi|^2+\alpha_n|\vec{\nabla}Q|^2+\frac{|\vec{\nabla}\times\vec{A}|^2}{8\pi}\right)d\vec{x}\nonumber\\
		&+&\iint_{\Omega}2\alpha_s\frac{\Lambda}{S_m}\left(|\hat{n}\cdot\vec{D}\psi|^2-\frac{|\vec{D}\psi|^2}{2}\right)d\vec{x}\label{energia libre usada}
	\end{eqnarray}
	where $\Omega \subseteq \mathbb{R}^2$ is the cross-section of the sample, $\Lambda$ is the geometrical coupling, $\alpha_s$ represents the superconducting stiffness, $\alpha_n$ denotes the scalar-tensor elasticity of the nematic molecules, $\psi\left(  \vec{x}\right)  =\rho\left(  \vec{x}\right) e^{i\theta\left(  \vec{x}\right)  }$ is the superconductor order parameter, $\ Q\left(  \vec{x}\right)  =S\left(
	\vec{x}\right)  e^{2i\alpha\left(  \vec{x}\right)  }$ is the nematic order parameter and $\vec{A}$ is the magnetic vector potential. In addition, 
	$\vec{D}=\vec{\nabla}-iq\vec{A}$ is the covariant derivative and $q$ is the effective charge of the model. Besides, $V \left(\psi(\vec{x}),Q(\Vec{x})\right)$ is the quartic potential given by 
	\begin{equation}\label{quartic_potential}
		V=-|a||\psi|^2+\frac{b|\psi|^4}{2}-|t||Q|^2+\frac{u|Q|^4}{2}+\nu|\psi|^2|Q|^2,
	\end{equation}
	being $\nu$ the coupling constant, and the other ones satisfy $|a|,|t|,b,u>0$ and $|\nu|^2\ll u b$ in the weak coupling condition. In this case, the minimizers for the quartic potential satisfy $\max|\psi|^2\simeq \rho_{m}^{2}:=\left\vert a\right\vert b^{-1}$ and $\max|Q|^2 \simeq S_{m}^{2} :=\left\vert t\right\vert u^{-1}$, which means that the mimimizers of the model are close to those of the uncoupled problem.
	The angle between the director $\hat{n}(\Vec{x})$ of the nematic molecules and the $\hat{z}$ axis is denoted by $\alpha(\vec{x})$ and the nematic vector field is given by $\hat{n}(\vec{x})=\sin\alpha\, \hat{x}+ \cos\alpha\, \hat{z}$. In this article, it is assumed that the connection between the nematic director and its environment is very small, accentuating our main objective on the interaction between nematicity and superconductivity itself. Introducing the normalized functions $f(\Vec{x})$ and $g(\Vec{x})$, we express the order parameters as $\psi=\rho_m f e^{i \theta}$ and $Q=S_m g e^{i 2\alpha}$, noticing that both unknowns are dimensionless and satisfy $0\leq |f|,|g| \leq 1$. Replacing both expressions into Eq.\eqref{energia libre usada}, the first-order conditions for the critical points of the energy are written as follows:
	
	
	\begin{align}
		\vec{\nabla}\times\vec{B}  & =4\pi\left[  (1-\Lambda g)\vec{J}_{s}+2 \Lambda g (\vec{J}_{s}\cdot\hat{n})\hat{n}\right]  =:4\pi\vec
		{J},\label{ley de ampere}\\
		0  & =-\alpha_{s}\ \vec{D}\cdot\left(  \left(  1-\Lambda g\right)
		\vec{D}(fe^{i\theta})+2\Lambda g\left(  \hat{n}\cdot\vec{D}(fe^{i\theta})\right)  \hat{n}\right) +\left( -|a| + \nu S_m^{2} g^2+b\rho_m^{2} f^2\right)f e^{i\theta},
		\label{GP superconductor}\\
		0  & =-\alpha_{n}\nabla^{2}g+4 \alpha_{n}g\left\vert \nabla\alpha\right\vert^{2}+2\alpha_{s}\Lambda\frac{\rho_m^2}{S_m^2}\left(  |\hat{n}\cdot\vec{D}(fe^{i\theta})|^{2}%
		-\frac{|\vec{D}(fe^{i\theta})|^{2}}{2}\right)  + \left(-|t| + \nu \rho_m^2 f^{2}+uS_m^{2} g^2 \right)g,%
		\label{GP nematico}\\
		0  & =-8\alpha_{n}S_m^{2}\rho_m^{-2}g^2\nabla^{2}\alpha+2\alpha_{s}\Lambda g\hat{n}_{\perp
		}\cdot\left((\hat{n}\cdot\vec{D}(fe^{i\theta}))^{\ast}\vec{D}(fe^{i\theta}) + (\hat{n}\cdot\vec
		{D}(fe^{i\theta}))\vec{D}(fe^{i\theta})^{\ast}\right]).\label{Ec nematica alpha}%
	\end{align}
	
	\noindent
	Here, the superconducting current $\vec{J}_s$ reads
	\begin{equation}\label{J_s}
		\vec{J}_s=2\alpha_s \rho_m^2 \left( q\text{Im}\left(fe^{-i\theta}\vec{\nabla}(fe^{i\theta})\right)- q^2 f^2\vec{A}\right)
	\end{equation}
	and the unitary vector $\hat{n}_{\perp}$ is given by:
	$\hat{n}_{\perp}=\cos\alpha\hat{x}-\sin\alpha\hat{z}.$ Since in this model it is assumed that the geometric coupling satisfies $\Lambda \ll 1$, it is natural to solve the problem by means of an iterative algorithm based upon an asymptotic expansion in powers of $\Lambda$. To do so, we shall henceforth consider $\Lambda=0$ and look for the first term in the expansion of each variable. Initially, using Eq. \eqref{GP superconductor} we get $\nabla \theta=0$, which implies that vorticity is necessarily neglected for the zero-order approximation. In the following we analyze two simple geometries in which the calculation is reduced to a 1-d system: an infinite slab and an infinite cylinder with a circular cross-section. In both cases, a uniform external magnetic field is applied to the sample. Particularly, for the slab case, we compare our results with those obtained in Refs.\cite{kwong-1995,aftalion_1997,brezis2005} 
	for the superconducting model, without vorticity.

	\subsection{Model derivation: The slab} \label{SecIIB} 
	In this Section we derive the equations that explain the response of an infinite slab of width $d$ under the influence of a uniform external magnetic field $\vec{B}(\vec{x})=H_{e}\hat{B}$, in which $\hat{B}$ is parallel to the boundary of the sample. Due to the symmetry with respect to translations in the direction of $\hat{B}$, the problem is mathematically equivalent to an infinite strip that is perpendicular to the magnetic field. Therefore, we may assume that the strip is parametrized as $0<x<d, z\in \mathbb{R}$ and also that the unknowns are written as: $\vec{A}=-A(x)\hat{z}$, $\psi(x)=\rho_m f(x) e^{i\theta(x)}$ and $Q(x)=S_m g(x) e^{2i\alpha(x)}$. Moreover,
	since we have set $\Lambda=0$, Eq. \eqref{GP superconductor} gives $\nabla \theta=0$, from which we deduce that $\psi$ is real (i.e., $\psi(x)=\rho_m f(x)$). Notice that $\vec{B}=\nabla \times \vec{A}= A^\prime(x) \widehat{y}$ and therefore $A^\prime$ represents the magnetic field inside the sample. The sought model for this configuration arises from setting the weak geometric coupling constant $\Lambda=0$ and replacing the unknowns in the equations of System \eqref{Sistema}. Before proceeding to the substitutions, we note that the third and fourth equations are scalar and real, while the first and second ones are vectorial in nature and will be expressed by their coordinates. Recalling the physical constants: 
	
	\begin{equation}\label{constantes fisicas}
		\begin{array}
			{ll}%
			\xi_{s}=\left( \frac{\alpha_{s}}{\left\vert a\right\vert} \right)^{1/2} & \text{sc. coherent length,} \\[1.25ex] 
			\xi_{n}=\left( \frac{\alpha_{n}}{\left\vert t\right\vert} \right)^{1/2}  & \text{nem. coherent length,} \\[1.25ex] 
			\kappa_{s}=\frac{\lambda_{L}}{\xi_{s}} & \text{sc. Landau-Abrikosov coeff.,}\\[1.25ex]
			\kappa_{n}=\frac{\lambda_{L}}{\xi_{n}} & \text{nem. Landau-Abrikosov coeff.,}\\[1.25ex]
			\lambda_{L}=\left( \frac{1}{8\pi q^{2}\alpha_{s}}\frac{b}{\left\vert a\right\vert} \right)^{1/2} & \text{penetration length,} \end{array}
	\end{equation}
	\noindent 
	introducing the scaling $x=\lambda_L x_a$, $A=q^{-1} \xi_{s}^{-1}A_a$ and setting $f_{a}(x_{a})=f(\lambda_L x_a)$, $g_{a}(x_{a})=g(\lambda_L x_{a})$, we get the following system (in which we omit the subscripts) posed in $0\leq x_{a} \leq d\lambda_L^{-1}$:  
	\begin{subequations}\label{sistema completo 1-d}
		\begin{align}
			0  & =-A^{\prime\prime}+f^{2}A\label{ec 2}\\
			0  & -\kappa_s^2 f^{\prime\prime} + f\left(A^2+f^2-1+\nu \frac{S_m^2}{|a|} g^2\right)   \label{ec 3}\\
			0  & =-\kappa_{n}^{-2}g^{\prime\prime}+g\left( 4\kappa_{n}^{-2}\left(  \alpha^{\prime}\right)  ^{2}+g^2-1+\nu \frac{\rho_m^2}{|t|}f^2\right)\label{ec 5}\\
			0  & =-f^{2}\alpha^{\prime\prime}.\label{ec 6}%
		\end{align}
	\end{subequations}
	
	As mentioned above, we are interested in capturing how the boundary conditions influence the solution, thus we set $f^{\prime}(0)=f^{\prime}(d\lambda_L^{-1})=0$, $g(0)=g(d\lambda_L^{-1})=0$, $A^{\prime}(0)=A^{\prime}(d\lambda_L^{-1})=H_e q \xi_s \lambda_L$ ($H_e$ is the external magnetic field). From Eq. \eqref{ec 6} we get $\alpha^{\prime}(x)=c_1$. Since we are looking for minimizers of the free energy, and $|\nabla Q|^2=S_m^2\left(g^{\prime\, 2}+4g^2\alpha^{\prime\,2}\right)$, we need to establish the possible values for $c_1$. Following a symmetry argument, we can assume that $\hat{n}(x=0)=\hat{n}(x=d\lambda_L^{-1})$, deducing that $c_1=\frac{k\pi \lambda_L}{d}$ with $k\in \mathbb{Z}$. Besides, we are interested in the existence of nontrivial solutions, hence $k=1$ and $\alpha^{\prime}=\frac{\pi \lambda_L}{d}$. We introduce the auxiliary parameters, expressed in terms of the physical constants, $B_e=H_e q \lambda_L^2 \kappa_s^{-1}$,
	\begin{equation*}\label{v y beta tilde}
		v=\dfrac{\nu\lambda_L^2}{(\alpha_s\,\alpha_n)^{1/2} S_m \rho_m \kappa_s\,\kappa_n}
		\text{ and } \;\;  \beta=\left(
		\dfrac{\alpha_n}{\alpha_s}\right)  ^{1/2}\dfrac{S_{m}\kappa_{n}}{\rho_{m} \kappa_{s}},    
	\end{equation*}
	in System \eqref{sistema completo 1-d}, and we obtain the following equations:
	\begin{equation*}
		\left\{
		\begin{array}{ll}
			A^{\prime\prime}=f^{2}A & \\[1.25ex]
			\kappa_{s}^{-2}f^{\prime\prime}=f\left(  A^{2}+f^{2}-1 + v \beta g^{2}\right) & \\[1.25ex]
			\kappa_{n}^{-2}g^{\prime\prime}=g\left(\frac
			{4\pi^2\xi_{n}^2}{d^2}+g^{2} -1+\dfrac{v}{\beta}f^{2}\right)& %
		\end{array}
		\right.
	\end{equation*}
	together with the boundary conditions $A^{\prime}(0)=A^{\prime}(d\lambda_L^{-1})=B_{e},\, f^{\prime}(0)=f^{\prime}(d\lambda_L^{-1})=0$ and $g(0)=g(d\lambda_L^{-1})=0.$ Finally, with the scaling $A=B_{e} A_1$ and dropping the subindex $1$ for $A,$ we get the equivalent system in which $B_{e}$ enters as a parameter:
	\begin{equation}\label{sistema simple 1-d}
		\left\{
		\begin{array}{ll}
			A^{\prime\prime}=f^{2}A & \\[1.25ex]
			\kappa_{s}^{-2}f^{\prime\prime}=f\left(  B_{e}^2 A^{2}+f^{2}-1+ v \beta g^{2}\right) & \\[1.25ex]
			\kappa_{n}^{-2}g^{\prime\prime}=g\left( \frac
			{4\pi^2\xi_{n}^2}{d^2}+g^{2} -1+\dfrac{v}{\beta}f^{2}\right)& %
		\end{array}
		\right.
	\end{equation}
	with the boundary conditions: $A^{\prime}(0)=A^{\prime}(d\lambda_L^{-1})=1,\, f^{\prime}(0)=f^{\prime}(d\lambda_L^{-1})=0$ and $g(0)=g(d\lambda_L^{-1})=0$. 
	We call System(\ref{sistema simple 1-d}) as the \textit{Nematic Ginzbug-Landau equation}. 
	Notice that, by setting $g\equiv 0$, we obtain the classical Ginzburg-Landau equation for the superconductor (see \cite{aftalion_1997,brezis2005}), whose solution shall be denoted $(A_{\text{sc}},f_{\text{sc}})$. We also note that by fixing $f\equiv 0$, the third equation of System \eqref{sistema simple 1-d} becomes the \textit{anharmonic} oscillator, whose nontrivial solution shall be denoted as $g_{\text{an}}$. 
	
	To close this subsection, we observe that although this scaling provides easy access to the uncoupled problems, the interaction term depends on two parameters, and this introduces the need to check whether the choice of parameters satisfy the weak coupling condition: $\max|\psi|^2\sim \rho_m^2$ and $\max|Q|^2\sim S_m^2$, where $\max|\psi|^2,\,\max|Q|^2$ are the critical points of the quartic potential \eqref{quartic_potential} and $\rho_m^2,\, S_m^2$ are those from the uncoupled problem (the problem without interaction), which ensures the scaling condition $|f|,|g|\leq 1$. In experiments, knowledge of the values of $\max|\psi|^2$, $\max|Q|^2$, $\rho_m^2$, $S_m^2$ and the Ginzburg-Landau parameters $\kappa_s$, $\kappa_n$, is enough to identify the magnitude of the interaction between both phases, denoted by $\nu$. Performing a simple calculation, it is easy to observe that the critical points of the quartic potential \eqref{quartic_potential} satisfy $\max|\psi|^2\leq \rho_m^2$ or $\max|Q|^2\leq S_m^2$ whenever $0<\nu\leq t\,\rho_m^{-2}$ or $0<\nu\leq a\,S_m^{-2}$ respectively, and this is the starting point to the test presented in Section \ref{IV}\,C (valid for $v>0$), in which this condition depends on the parameters $v$ and $\beta$. On the other hand, for $v<0$, the situation is completely different, since the same computation shows that for any $\nu<0$, the scalings $\psi=\rho_m f$ and $Q=S_m g$ satisfy $\max|f|,\max|g|\leq 1$ only for specific values of $|v|$ and $\beta$, which are related with the appearance of \textit{supersaturation}, see Section \ref{IV}\,A\,2.

	\subsection{Model derivation: the cylinder} \label{SecIIC} 
	
	In this Section we study the case where the nematic superconductor medium is a cylinder with cross-section $\Omega=\{0\leq r\leq R \}$.  We consider the effect of a uniform external magnetic field normal to $\Omega$. Since we are looking for radial solutions, it is natural to set the ansatz $\psi=\rho_m f(r) e^{i\phi}$, $Q=S_m g(r) e^{i2\phi}$ and $\vec{A}=A(r)\hat{\phi}$. Setting $\Lambda=0$, introducing the scaling $s=r \lambda_L^{-1}$, replacing $\Vec{A},f,g$ into Eq.\eqref{Sistema} and performing the Gauge transformation $A\leftarrow A-\frac{1}{qr}$, we get the system posed for $s\in [0,R\lambda_L^{-1}]$:
	\begin{equation}\label{sistema polares simplificado}
		\left\{
		\begin{array}
			[c]{lcl}%
			A^{\prime\prime}+\frac{1}{s}A^{\prime}-\frac{1}{s^{2}}A=f^{2}A & &\\[1.25ex]
			\kappa_{s}^{-2}\left(  f^{\prime\prime}+\frac{1}{s}f^{\prime}\right)  =f\left(
			B_{e}^2A^{2}+f^2-1+ v \beta g^{2}\right) & & \\[1.25ex]
			\kappa_{n}^{-2}\left(  g^{\prime\prime}+\frac{1}{s}g^{\prime}-\frac{4}{s^{2}%
			}g\right)  =g(g^2+\frac{v}{\beta}f^{2}-1) & & %
		\end{array}
		\right.
	\end{equation}
	with the boundary conditions $f^{\prime}(R\lambda_L^{-1})=0$, $g(R\lambda_L^{-1})=0$ and $A^{\prime}(R\lambda_L^{-1})=1$. We also need to add regularity conditions at the origin: $f^{\prime}(0)=0$, $A(0)=0$, $A^{\prime}(0)=0$ and $g(0)=0$, which we take for granted since the left side of the System \eqref{sistema polares simplificado} represents a Bessel equation of order 0, 1, 2, respectively. It is worth mentioning that previously in Ref.\cite{DeLeo2022}, numerical studies have been performed in a system equivalent to \eqref{sistema polares simplificado} but with Dirichlet or Neumann boundary conditions separately, while in this work we consider Neumann boundary conditions for $f(r)$ and Dirichlet conditions for $g(r),$ as stated 
	
	\section{Theoretical approach: Fr\'{e}edericksz threshold} \label{III}
	
	As is expected for nematic samples, the existence of a nontrivial response depends upon certain threshold conditions on the nematic order parameter. This Section is devoted to showing that for the slab and the cylinder cases, there exist critical values below for which the response is trivial; these thresholds have a geometrical nature and are expressed in terms of the coherent length $\xi_n$ and the characteristic dimensions of the sample. Regarding the superconducting parameter, equivalent results imply the existence of critical values for the external magnetic fields, see \cite{Tinkham}. This analysis is deferred to Section  \ref{IV}, where we show that the presence of nematicity reduces the upper critical value for the magnetic field. 
	
	\subsection{Threshold in the slab}
	
	Aiming to obtain the threshold condition, we multiply the third equation of System \eqref{sistema simple 1-d} by $g$, we integrate by parts on $[0,d\lambda_{L}^{-1}]$ and we get the identity
	\begin{align*}
		\left(  1-4\pi^{2}\frac{\xi_{n}^{2}}{d^{2}}\right)  \left\Vert g\right\Vert
		_{L^{2}[0,d\lambda_{L}^{-1}]  }^{2}   =\kappa_{n}^{-2}\left\Vert
		g^{\prime}\right\Vert _{L^{2}[0,d\lambda_{L}^{-1}]}^{2}\\
		+\dfrac
		{v}{\beta}\left\Vert fg\right\Vert _{L^{2}[0,d\lambda_{L}^{-1}]  }^{2}+\left\Vert g\right\Vert _{L^{4}[0,d\lambda_{L}^{-1}]  }^{4}\,.
	\end{align*}
	
	\noindent 
	For $v>0$, we have the estimate
	\[
	\left(  1-4\pi^{2}\frac{\xi_{n}^{2}}{d^{2}}\right)  \left\Vert g\right\Vert
	_{L^{2}[0,d\lambda_{L}^{-1}]}^{2} \geq \kappa_{n}^{-2}\left\Vert g^{\prime}\right\Vert
	_{L^{2}[0,d\lambda_{L}^{-1}]}^{2}\,.
	\]
	On the other hand, for $v<0$ we have the estimate
	\[
	\left(  1-4\pi^{2}\frac{\xi_{n}^{2}}{d^{2}}\right)  \left\Vert g\right\Vert
	_{L^{2}}^{2} 
	+\dfrac
	{|v|}{\beta}\left\Vert g\right\Vert _{L^{2}}^{2}
	\geq \kappa_{n}^{-2}\left\Vert g^{\prime}\right\Vert
	_{L^{2}}^{2}\,,
	\]
	\noindent 
	in which the $L^2$ norm is understood in $[0,d\lambda_L^{-1}]$. Since the first eigenvalue of the laplacian on $[0,d\lambda_{L}^{-1}]$ with Dirichlet boundary conditions is $\lambda_1=\pi^2 \lambda^2_L d^{-2}$, we deduce that 
	\[
	\left\Vert g^{\prime}\right\Vert _{L^{2}[0,d\lambda_{L}^{-1}] }^{2}\geq \pi^{2}\lambda_{L}^{2}d^{-2}\left\Vert g\right\Vert
	_{L^{2}[0,d\lambda_{L}^{-1}] }^{2}\,.
	\]
	\noindent 
	Combining previous estimates for $v>0$, we get the inequality
	\[
	\left(  1-4\pi^{2}\frac{\xi_{n}^{2}}{d^{2}}\right)  \Vert g\Vert_{L^{2}[0,d\lambda_{L}^{-1}] }%
	^{2}\geq \kappa_{n}^{-2}\lambda_{L}^{2}\pi^{2}d^{-2}\Vert g\Vert_{L^{2}[0,d\lambda_{L}^{-1}] }^{2}\,.
	\]
	In addition, since $\kappa_{n}=\lambda_{L}/\xi_{n},$ we infer that the following upper bound for the threshold condition must be satisfied to get a nontrivial response:
	\begin{equation}\label{umbral 1-d}
		\dfrac{d}{\xi_{n}}\geq \sqrt{5}\pi=7.02481\cdots\,,
	\end{equation}
	which establishes a relationship between the nematic coherent length $\xi_n$ and the strip width $d$, independently of $v$ and $\beta$. In contrast, performing analogous calculations for $v<0$, we deduce a threshold value that depends upon $v$ and $\beta$: 
	\begin{equation*}
		\dfrac{d}{\xi_{n}}\geq \sqrt{5}\,\pi \left(1+\frac{|v|}{\beta} \right)^{-1}
	\end{equation*}
	
	\subsection{Threshold in the cylinder}
	
	Regarding the case of the cylinder, we proceed in a similar fashion to what has been done for the slab. The main difference is related to the presence of the Bessel operator of order $\eta=2$ (with Dirichlet boundary conditions). Let us recall the self-adjoint expression for the Bessel operator of order 2: $L_2(g(s))=-(sg^{\prime}(s))^{\prime}+\frac{2^2}{s} g(s)$. We also recall the basic estimate provided by Rayleigh quotient:
	\begin{equation*}
		\langle L_2 g,g\rangle \geq \lambda_1 \int_{[0,R\lambda_{L}^{-1}] } |g(s)|^2 s ds=:\|g\|^2_{L^2_w[0,R\lambda_{L}^{-1}] }
	\end{equation*}
	in which $w(s)=s$ is the weight function and $\lambda_1=5.13562\cdots$ is the first non-zero root for $J_2$ (the Bessel function of order 2). Multiplying the third equation of System \eqref{sistema polares simplificado} for $s\, g(s)$, integrating by parts, rearranging terms and assuming $v>0$, we have
	\[
	\left\Vert g\right\Vert _{L_{w}^{2}[0,R\lambda_{L}^{-1}]}^{2}%
	\geq \kappa_{n}^{-2}\left\langle L_2g,g\right\rangle\,. 
	\]
	
	\noindent 
	Taking into account Rayleigh quotient we get
	\[
	\left\Vert g\right\Vert
	_{L_{w}^{2}\left[  0,R\lambda_{L}^{-1}\right]  }^{2}
	\geq \lambda_1 \kappa_{n}^{-2} \left\Vert g\right\Vert
	_{L_{w}^{2}\left[  0,R\lambda_{L}^{-1}\right]  }^{2}\,,
	\]
	from where we deduce the following upper bound for the  Fr\'{e}edericksz threshold in the disk, for $v>0$, 
	\[
	\frac{R}{\xi_{n}}\geq \sqrt{\lambda_1}=2.26619\cdots\,,
	\]
	
	\noindent 
	in which we have used that $\kappa_n=\frac{\lambda_L}{\xi_n}$. For $v<0$, we have an approximation that depends upon $|v|$ and $\beta$:
	\begin{equation*}
		\frac{R}{\xi_{n}}\geq \sqrt{\lambda_1}\left(1+\frac{|v|}{\beta} \right)^{-1}\,.
	\end{equation*}

	\section{Numerical Results} \label{IV}
	We complement the previous calculations with a numerical analysis of the \textit{Nematic Ginzburg-Landau equation} in the case of the slab [System \eqref{sistema simple 1-d}], showing the existence of symmetric nematic-superconducting solutions, that we tagged as \textit{mixed solutions}. In this case, they are even with respect to its center $x_m:=d(2\lambda_L)^{-1}$, for different values of $v$ and $\beta$. 
	Our computations show that the region in the $(\beta,\, v)$ plane containing \textit{mixed solution} is small enough to make it difficult to find the correct regime of parameters. Our results were obtained working with $d$ and $\xi_n$ such that $d\cdot \xi_n^{-1}=20$, which satisfies the threshold condition \eqref{umbral 1-d}, and $\kappa_s=\kappa_n=5$, which corresponds to type II superconductors. We solve System \eqref{sistema simple 1-d} by applying the Procedure \ref{alg:main_solver} detailed below.
	\medskip
	
	\begin{algorithm}
		\caption{Main Solver}
		\label{alg:main_solver}
		\begin{algorithmic}
			\Require $f_0$ 
			\Comment{A distribution in $[0,x_m]$}
			\Require \texttt{Tol} 
			\Comment{Tolerance}
			\State $N\gets 0$
			\State $\texttt{Err}_N \gets 1+\texttt{Tol}$
			\While{$\texttt{Err}_N \geq \texttt{Tol}$}
			\State $g_N\gets \texttt{SolveEq3}[g(x_m)=\tau ,g^{\prime}(x_m)=0]$
			\Statex
			\Comment{Take $f=f_N$}
			\Statex 
			\Comment{$\tau$: shooting parameter}
			\State $A_N\gets \texttt{SolveEq1}[A(x_m)=0,A^{\prime}(x_m)=\tau]$
			\Statex
			\Comment{Take $f=f_N$}
			\Statex 
			\Comment{$\tau$: shooting parameter}
			\State $f_{N+1}\gets \texttt{SolveEq2}[f(x_m)=\tau,f^{\prime}(x_m)=0]$
			\Statex
			\Comment{Take $(g,A)=(g_N,A_N)$}
			\Statex 
			\Comment{$\tau$: shooting parameter}
			\State $\texttt{Err}_{N+1}\gets \max\left|f_{N+1}-f_N\right|/f_N(x_m)$
			\Statex
			\Comment{Relative error}
			\State $\texttt{Rate} \gets\frac{\texttt{Err}_{N+1}}{\texttt{Err}_N}$
			\If{$\texttt{Rate}\geq 0.85$}
			\State \textbf{Exit}
			\Comment{Convergence is not guaranteed}
			\EndIf
			\State $N\gets N+1$
			\EndWhile
			\Ensure $(A,g,f)\gets (A_N,g_N,f_N)$ 
			\Statex
			\Comment{Approximate solution}
		\end{algorithmic}
	\end{algorithm}
	
	Notice that each time the solver is called, $\texttt{SolveEq}\#$ solves the corresponding evolution equation, backward in time, in the interval $[0, x_m]$ by means of the Shooting Method (see \cite{SB} for details). The shooting parameter in each case is given by $ \tau=f(x_m)$, $\tau=g(x_m)$ and $\tau=A^\prime(x_m)$. Besides, the target is given by $f^\prime(0)=0$, $g(0)=0$ and $A^\prime(0)=1$, respectively. The figures depicted below are developed accordingly by setting $\texttt{Tol}:=10^{-4}$. We remark that in all cases the computed value for \texttt{Rate} is always below $0.5$, ensuring the linear convergence of the algorithm. Finally, we note that since in Procedure \ref{alg:main_solver} both $g_N$ and $A_N$ depend on $f_N$, the convergence of $f_N$ ensures the convergence of $(A_N,g_N,f_N)$. \bigskip 
	
	Following \cite{Tinkham}, we recall that for type II superconductors there are two critical values of the external magnetic field $B_e$, called $H_{c1}$ and $H_{c2}$, that establish three different regions: (a) for $B_e>H_{c2}$, the response is metallic, which means $f\equiv 0$; (b) for $B_e$ in $[H_{c1},H_{c2}]$, the response is mixed metallic-superconducting, which means $0\leq f(x) \leq 1$; (c) for $B_e<H_{c1}$, the response is pure superconducting, which means $f\equiv 1$. 
	In this paper, we are working with $\kappa_s=\kappa_n=5$, which corresponds to a type II superconductor, and we explore how the presence of a nematic phase influences the critical magnetic fields. Since, for $v>0$, both nematicity and superconductivity compete for the energy, it is natural to expect that the presence of nematicity decreases the critical value for $B_e$ needed to destroy superconductivity. \bigskip 
	
\begin{figure}[ht]
		\centering
		\includegraphics[height=6cm, width=8cm]{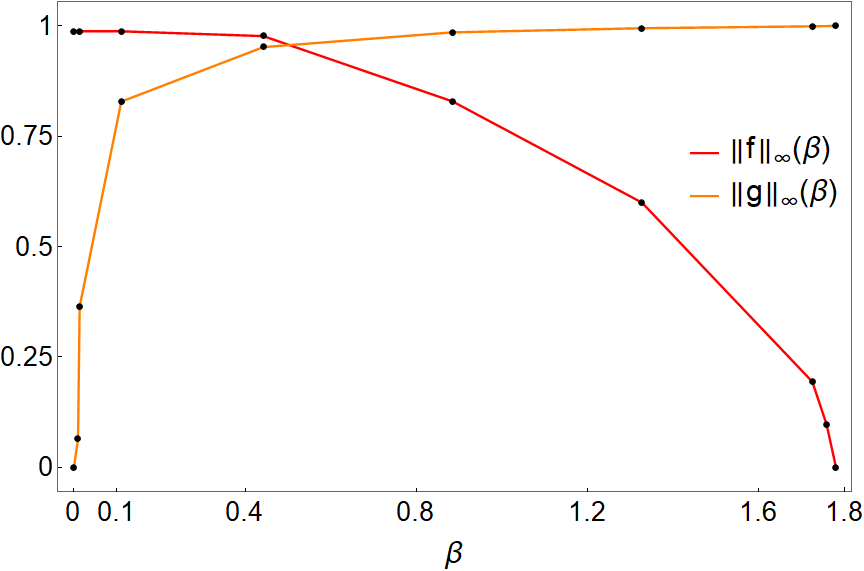}
		\caption{(Color online) For $B_e=1$ and $v=0.5$, continuous transition from the superconducting state ($f\equiv f_{\text{sc}}$ and $g\equiv 0$) to the nematic state ($g\equiv g_{\text{an}}$ and $f\equiv 0$), as $\beta$ varies from $\beta\sim 0.015$ to $\beta \sim 1.76$. }
		\label{fig:MaxFG}
	\end{figure}

	Our computations show, as expected, that for each fixed $(v,\beta)$, there are new critical values, henceforth denoted by $H_{c1}^{*}(v,\beta)$ and $H_{c2}^{*}(v,\beta)$, such that if $B_e\leq H_{c1}^{*}(v,\beta)$, the system has a superconducting response ($g\equiv  0$),  and if $B_e\geq H_{c2}^{*}(v,\beta)$, the response is strictly metallic ($f\equiv 0$).
	In addition, we have computed the critical values for $\kappa_s=\kappa_n=5$ and $d\cdot \xi_n^{-1}=20$, and we have obtained: $H_{c1}=0$ and $H_{c2}=4.99999$. With this in mind we have applied Procedure 1 for fixed values of $B_e$ and for special choices of the parameters $(v,\beta)$, and we have checked the existence of \textit{mixed solutions}. More precisely, fixing the value of $B_e$ and $v$, we have found an interval for the $\beta$ parameter that produces these \textit{mixed solutions}. This confirms the existence of a bounded interval for the $\beta$ parameter (depending only on $B_e$ and $v$) within which the continuous transition from the superconducting state $g\equiv 0$ to the nematic state $f\equiv 0$ takes place. In Fig.\ref{fig:MaxFG}, we plot the maximum value of $f$ and $g$ as a function of $\beta$, for $B_e=1$  and $v=0.5$; this picture shows that lowering the value for $\beta$ below $0.008$, we get the superconducting state $f\equiv f_{\text{sc}}$, $g\equiv 0$, and raising it above $1.76$, we get the nematic state $g\equiv g_{\text{an}}$, $f\equiv 0$. In terms of the critical values already introduced, the results shown in Fig.\ref{fig:MaxFG} should be interpreted as $H_{c2}^*(0.5, 1.76)=B_e=1$. On the other hand, for $\beta<0.008$, the solution $f\equiv f_{\text{sc}}$, $g\equiv 0$, is not related with $H_{c1}$, since $f_{\text{sc}}$ is far from being a \textit{pure superconductor} state, see Fig.\ref{tres_dibujos_posit}, panel (a). In this section we present three different results: we start working with $B_e=1$ and $v=0.5$, and we present \textit{mixed solutions} that correspond to the transitions $g\sim 0$ and $f\sim 0$, see Fig.\ref{tres_dibujos_posit}, panels (a) and (b), respectively; together with a \textit{mixed solution} which is far from transition, see Fig.\ref{tres_dibujos_posit}, panel (c). Fig.\ref{tres_dibujos_posit} as a whole, establishes an interval that contains the mixed solutions, which is strictly contained in $[0.008,1.8]$. We also consider $B_e=1$ and $v=-0.02$, where $f$ and $g$ interact in collaboration. However, this framework is totally different, since the existence of an interval containing mixed solutions is not a consequence of the transitions (from superconducting to mixed, and from mixed to nematic) as for $v>0$, but for the non-compliance with the weak coupling condition, expressed by the supersaturation effect, see details below on Section \ref{IV} A 2. Fig.\ref{tres_dibujos_neg}, panels (a) and (b), establishes the existence of two critical values $0.1\leq \beta$ and $\beta \leq 1.4$ between which there are mixed solutions, but for $\beta\leq 0.1$ or $\beta\geq 1.4$, the solution satisfies $1<\max |g|$ or $1<\max |f|$ respectively, and this indicates that the weak coupling condition is not satisfied. Next, we fix $v=0.5$ and $\beta=0.8$, and we consider $B_e=3< 4.99999$, slightly below the computed value for the critical field $H_{c2}$, trying to capture the critical value $H_{c2}^*(0.5,0.8)$ needed to cancel the superconductivity. We conclude this section by proposing a practical criterion for checking whether the chosen parameters give rise to a model that satisfies the weak coupling condition. 
	
	\subsection{Existence of mixed states for $B_e=1$}
	Firstly, let us take $B_e=1$ to prove that under particular choices of $v$ and $\beta$, the corresponding solutions reveal a different interplay between nematicity and superconductivity. According to Fig.\ref{fig:MaxFG}, we fix $v$ and consider different values of $\beta$ looking for the transition values reported there.
	
	\subsubsection{Case $v>0$: $f$ and $g$ interact in competition}
	
\begin{figure*}[ht]
		\centering
		\includegraphics[width=5cm]{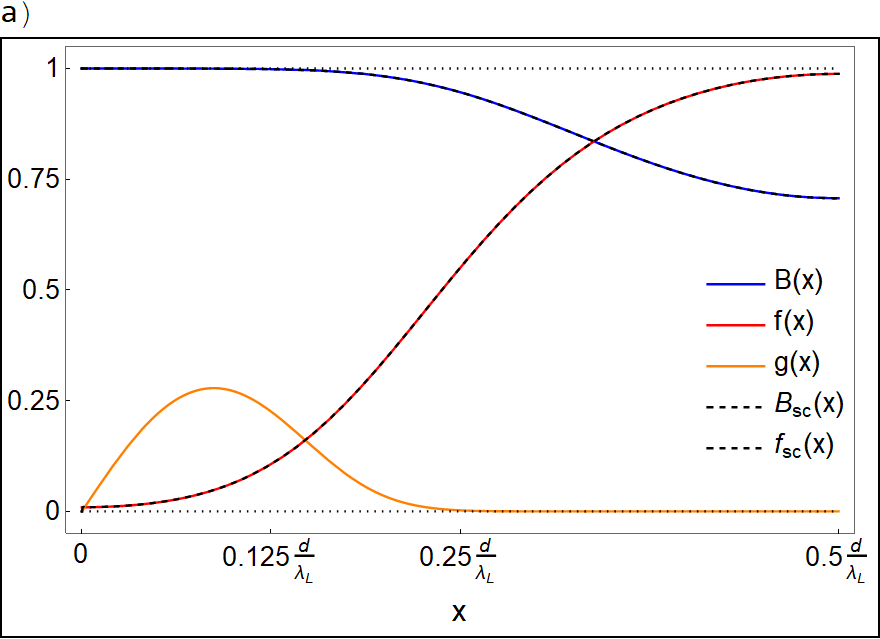}
		\hspace{.5cm}
		\includegraphics[width=5cm]{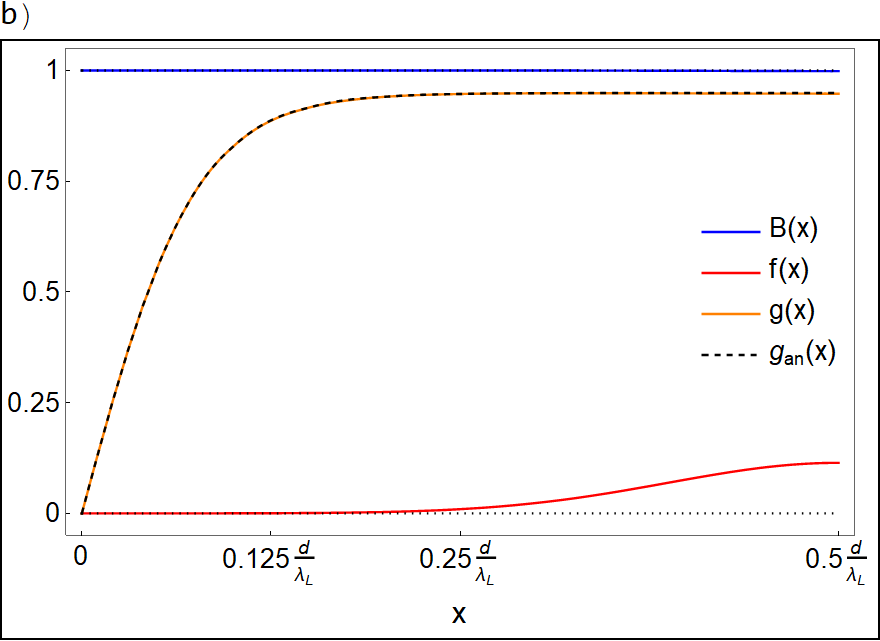}
		\hspace{.5cm}
		\includegraphics[width=5cm]{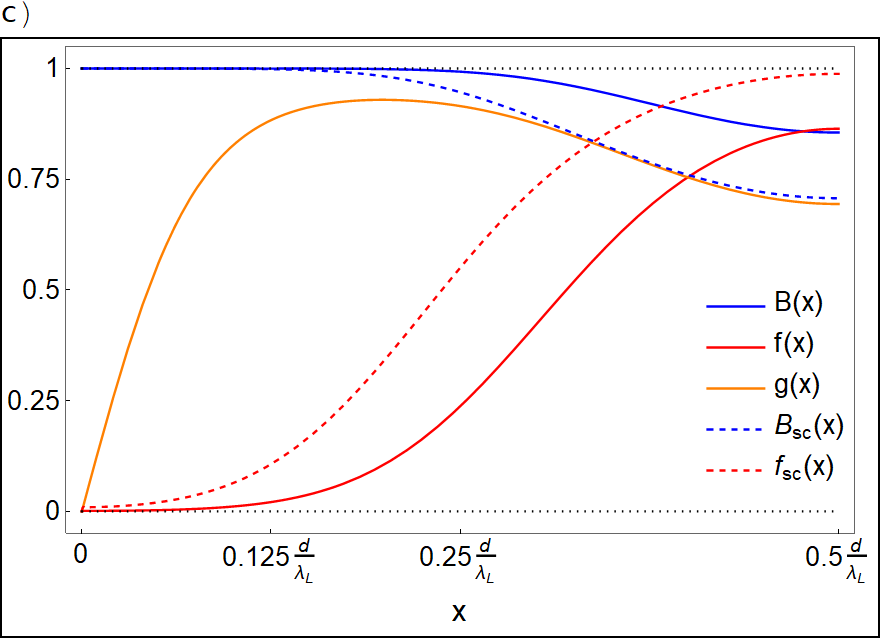}
		\caption{(Color online) Magnetic field $B(x)$ (solid blue line), superconductor order parameter $f(x)$ (solid red line), and nematic order parameter $g(x)$ (solid yellow line) as a function of the distance in the strip with respect to the origin for $B_e=1$ and $v=0.5$, in the situations a) $\beta=0.015$ (near-to-superconducting response), b) $\beta=1.76$ (near-to-nematic state) and c) $\beta=0.8$ (nematic-superconducting state). As a reference, we add the corresponding superconducting profiles $B_{\text{sc}}(x)$ and $f_{\text{sc}}(x)$, and the anharmonic solution to the nematic order parameter $g_{\text{an}}(x)$ in dashed lines, whenever is needed.}
		\label{tres_dibujos_posit}
	\end{figure*}

	Assuming $v>0$, we can prove that the order parameters $f(x)$ and $g(x)$ interact in competition. Also, we demonstrate that for each fixed $v>0$ there is a bounded interval in $\beta$ such that the sample has a nontrivial response. Our findings for the profiles of the order parameters and the magnetic field, fixing $v=0.5$, and for the lower and the upper limits in the  $\beta$ range are reported in Figs. \ref{tres_dibujos_posit}(a) and \ref{tres_dibujos_posit}(b). In panel (a), corresponding to $\beta=0.015$, we have $g(x)\leq 0.275$ that tends to vanish as $x\to 0.5 d\lambda_L^{-1}$. Moreover, $f(x)$ and $B(x)$ converge to the solution of the superconducting case $f_\text{sc}(x)$ and $B_\text{sc}(x)$ in the same limit. Actually, the exponential decrease in the magnetic field is of the same magnitude as the maximum of $g(x)$. Furthermore, whereas the magnetic field reaches its minimum in the bulk, the nematic order parameter accommodates its maximum close to the boundary. On the other hand, In panel (b) corresponding to $\beta=1.76,$ we obtain $f(x)\simeq 0$, with a small magnitude when $x\to 0.5d \lambda_L^{-1}$, $B(x)\simeq 1$ and $g(x)\simeq g_\text{an}(x)$, where $g_\text{an}$ is the solution of the anharmonic-like equation $g^{\prime\prime}=\kappa^2 \left( \frac{4\pi^2 \xi_n^2}{ d^2}+g^2-1\right) g$, $g(0)=0$, $g^{\prime}(x_m)=0$. For $\beta>1.8$, we obtain the trivial solution $f(x)\equiv 0$, $B(x)\equiv 1$ and $g(x)\equiv g_\text{an}(x)$, where the nematic density and the magnetic field inside the strip are maximized. For the sake of completeness, if we take the same value of $v$ and an intermediate value of $\beta$, say,  $\beta=0.8$, we sketch the corresponding profiles in Fig.\ref{tres_dibujos_posit}(c). Here, the nematic profile increases as we go inside the strip, lowering the value of $f(x)$ in comparison with the pure superconducting case. Then, the magnetic field in the strip tends to penetrate further into the interior of the region. For all these reasons, it is clear that $f(x)$ and $g(x)$ compete with each other.
	
	\subsubsection{Case $v<0:$  $f$ and $g$ interact in collaboration }
\begin{figure*}[ht]
		\centering
		\includegraphics[width=5cm]{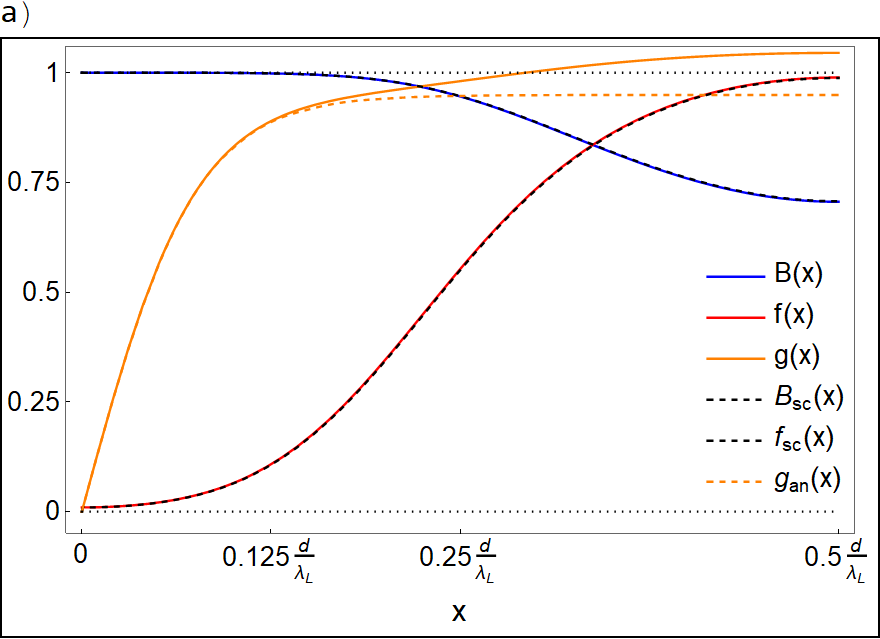}
		\hspace{.5cm}
		\includegraphics[width=5cm]{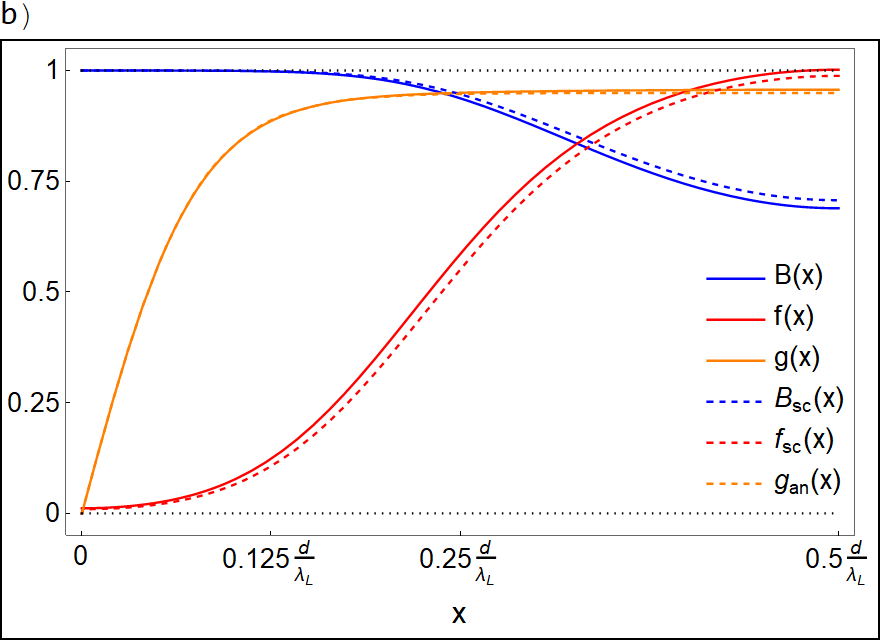}
		\hspace{.5cm}
		\includegraphics[width=5cm]{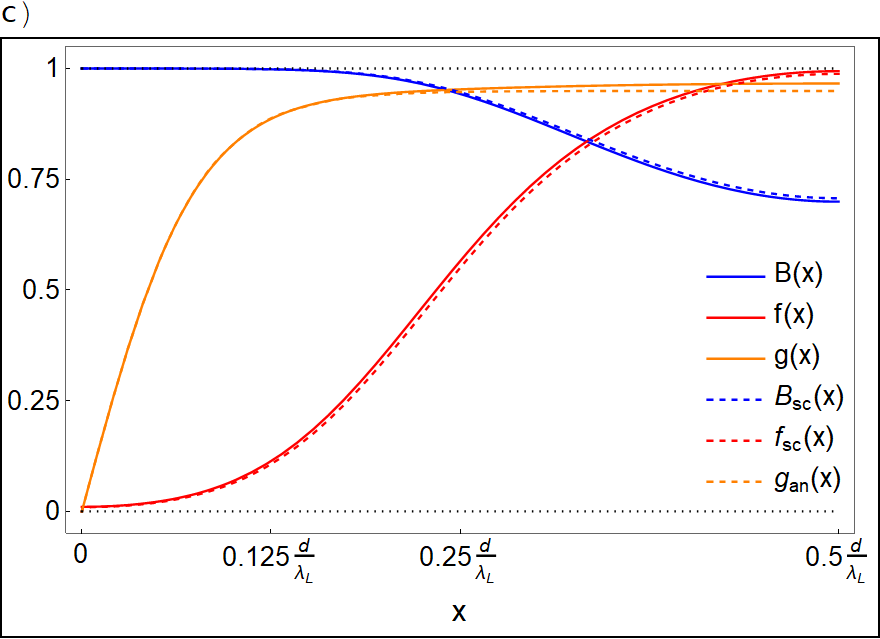}
		\caption{(Color online) Magnetic field $B(x)$ (solid blue line), superconductor order parameter $f(x)$ (solid red line), and nematic order parameter $g(x)$ (solid yellow line) as a function of the distance in the strip with respect to the origin for $B_e=1$ and $v=-0.02$, in the situations a) $\beta=0.1$, b) $\beta=1.4$ and c) $\beta=0.6$. As a reference, we add the corresponding superconducting profiles $B_{\text{sc}}(x)$ and $f_{\text{sc}}(x)$, and the anharmonic solution to the nematic order parameter $g_{\text{an}}(x)$ in dashed lines, whenever is needed.}
		\label{tres_dibujos_neg}
	\end{figure*}

	If we now consider the case $v<0$, we can verify a collaborative relationship between both order parameters $f(x)$ and $g(x)$. This behavior is expected because $v$ in the quartic potential term of the original Helmholtz Free Energy given by Eq.\eqref{energia libre usada} is the crucial factor describing the interaction between the nematic and superconducting components. As it was noted previously, for $v<0$, the situation is completely different, since the scaling conditions $|f|,|g|\leq 1$ hold for specific values of parameters $v$ and $\beta$. To this end, we set $v=-0.02$, a \textit{small} value for $v$, and explore the response for different values of $\beta$ trying to capture the existence of mixed solutions. The profiles for the magnetic field, superconducting, and nematic order parameters of the model, for $\beta=0.15$, $1.4$ and $0.6$, are illustrated in Figs.\ref{tres_dibujos_neg}(a-c), respectively. If 
	$\beta=0.15,$ [Fig.\ref{tres_dibujos_neg}(a)], the profile for the nematic parameter verifies $g(x_m)>1$, which indicates that the weak coupling condition does not hold for the choice $(v=-0.02,\beta=0.15)$; we shall call this behavior as \textit{nematic supersaturation}. For $\beta=1.4$ [Fig.\ref{tres_dibujos_neg} (b)], we can observe the presence of \textit{superconducting supersaturation} for $f(x)$: $f(x_m)>1$ (the actual value is $f(x_m)=1.0018\ldots$), which lead us to conclude that the weak coupling condition does not hold for $(v=-0.02, \beta=1.4)$. At intermediate values of $\beta$, say, $\beta=0.6$ [ Fig.\ref{tres_dibujos_neg}(c)], we can identify a collaborative state between densities $f(x)$ and $g(x)$, in which the presence of one reinforces the presence of the other. Besides, in this situation, the response of $f(x)$, $g(x)$ and $B(x)$ are close to the solution of the uncoupled system given by $f_{\text{sc}}(x)$, $g_{\text{an}}(x)$ and $B_{\text{sc}}$ (dashed lines in Fig.\ref{tres_dibujos_neg}); however, in panel c) it is possible to detect a small difference among the related components; in addition, notice that $B(x)<B_{\text{sc}}(x)$ and $f(x)>f_{\text{sc}}(x)$, which means that the presence of nematicity reinforces the superconducting response, and since $g(x)>g_{\text{an}}(x)$, this is a clear evidence of a collaborative interaction. We also made numerical experiments with $v\leq -0.05$,  and in all cases our numerical solutions are totally supersaturated: which indicates that the weak coupling condition is not satisfied for this range of parameters.
	
	\subsection{Existence of mixed states for $B_e\sim 0$ and $B_e=3$}
	
	We now move forward to the exploration of the response of the system in Eq.\eqref{sistema simple 1-d}, when the magnitude of the external magnetic field $B_e$ is near $0$, and the case when $B_e$ is greater than $1$ but under the critical value $H_{c2}$. In the frame of nematic-superconducting materials, we expect a variation of the spatial profile for the nematic and superconductor order parameters, and of the internal magnetic field, as long as we modify $B_e$. In addition, we will focus on the case $v>0$ because we are interested in the competition behavior of the order parameter profiles and their physical consequences. Then, we set $v=0.5$ and $\beta=0.8$ in order to compare with the previous results. Moreover, as it was mentioned earlier and due to the competitive nature of the interaction, we expect that the nematic condition of the material produces a new upper critical value depending on $(v,\beta)$, which we have denoted by $H_{c2}^{*}(v,\beta)$, that is smaller than the upper critical value $H_{c2}$ (we recall that $H_{c2}$ is the minimum value of $B_e$ needed to destroy superconductivity); in other words, we expect to obtain $H_{c2}^{*}(v,\beta)\leq H_{c2}$ for any choice of $v$, $\beta>0$. \bigskip    
	
\begin{figure*}[ht]
		\centering
		\includegraphics[width=6cm]{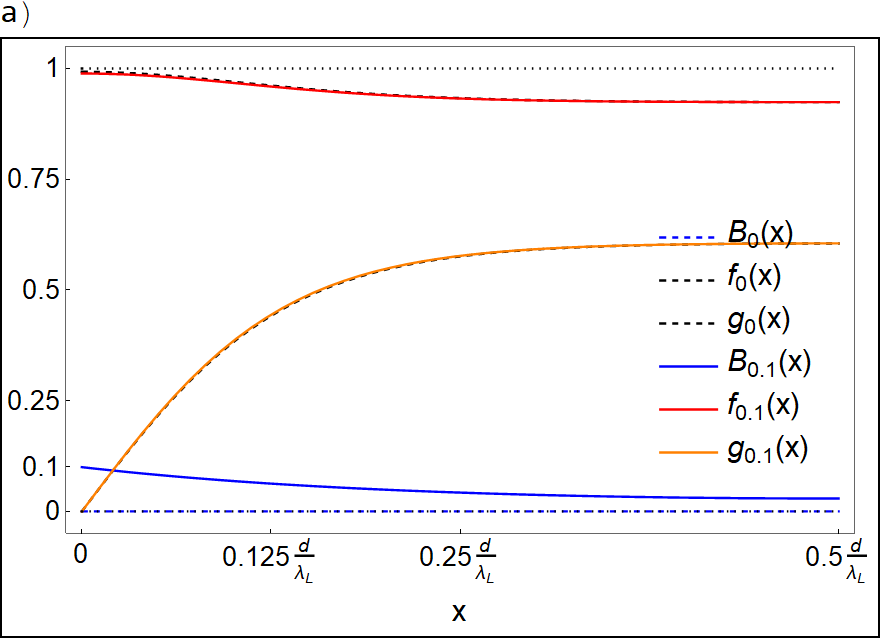}
		\hspace{1cm}
		\includegraphics[width=6cm]{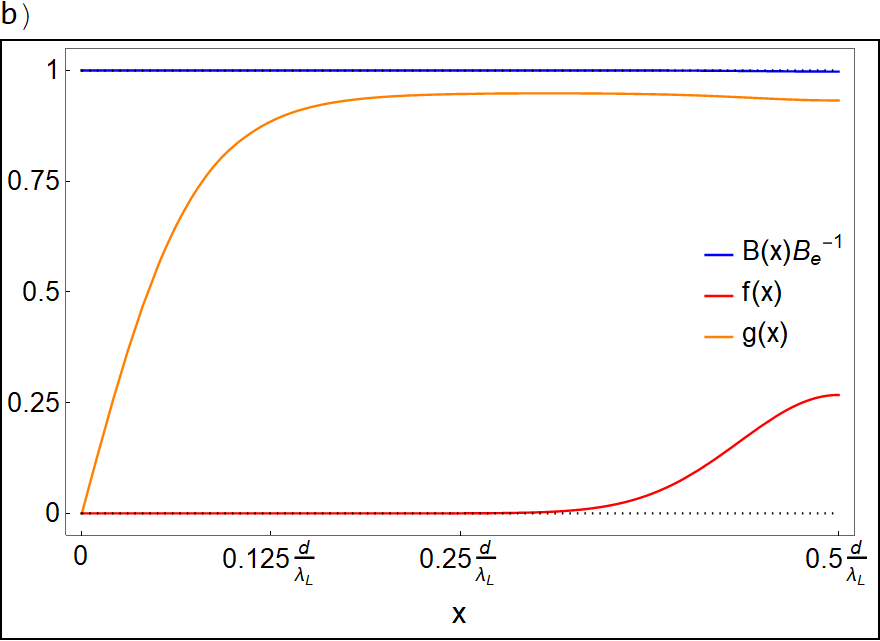}
			\caption{(Color online) Magnetic field $B(x)$ (blue line), superconducting order parameter $f(x)$ (red line), and nematic order parameter $g(x)$ (yellow line) as a function of the distance in the strip with respect to the origin. We fix the parameters as $v=0.5$ and $\beta=0.8$ for the applied magnetic fields a) $B_e=0$ (dashed lines) $B_e=0.1$ (solid lines) and (b) $B_e=3$.}
		\label{fig:Bcrit}
	\end{figure*}

	Let us begin with the case $B_e\sim 0$, with $(v,\, \beta)=(0.5,\, 0.8)$, plotted in Fig.\ref{fig:Bcrit}(a). As we can deduce from it, although the superconductor order parameter satisfies $f(x)\sim 1$,  the nematic one $g(x)$ is far from being small.  In this case, we observe that the small value of the magnetic field allows the nematic order parameter to increase inside the sample, which explains the small but observable variation of the superconducting order parameter.\bigskip 
	
	Next, we analyze the case $1<B_e <H_{c2}$ [see Fig. \ref{fig:Bcrit}(b)]. As it was mentioned previously, for $\kappa_s=5$ we have obtained $H_{c2}= 4.99999$. Throughout the exploration of the instance $v=0.5$ and $\beta=0.8$, for different values of $B_e$, we have obtained for $B_e=3$ a solution with $f(x)\sim 0$,  from where we deduce that System \eqref{sistema simple 1-d} is near the critical value and thus $3<H_{c2}^{*}(0.5,\, 0.8)$. In addition, numerical computations with $B_{e}=4$ (not reported here) yield $f(x)\equiv 0$, implying that $3<H_{c2}^{*}(0.5,\, 0.8)<4$. This is in coincidence with the inequality $H_{c2}^{*}(v,\, \beta) \leq H_{c2}$ indicated above. 
	
	\subsection{Consistency test of the weak coupling condition} \label{subsect:IV_C}
	We close this Section by offering a criterion to verify whether the set of parameters involved in the numerical computations are consistent with a model that satisfies the weak coupling condition. According to the results presented in Section \ref{IV}\,A, we focus on the case $v>0$, in which $f$ and $g$ interact in competition.  
	In this case, it is expected that if $f$ prevails over $g$, then the superconducting phase will dominate the energy and there will be no energy to align the nematic molecules. Now, we assume that there is a critical value at which $g$ starts to appear. Analogously, we can assume the existence of another critical value related to the appearance of $f$. The idea is to consider both scenarios as the \textit{zero-coupling} situation and then use these critical values as a reference to estimate the magnitudes for which the transitions $g\sim 0$ and $f\sim 0$ occur. We look at the equations of the system \eqref{sistema simple 1-d} for $v>0$ and we introduce the quantities $\gamma_G:=v \beta^{-1}$ and $\gamma_F:=v \beta$ (the coupling constant in the equation of $g$ and $f$, respectively). For $\beta\to 0$, we have $\gamma_F\to 0$, implying that the value of $g$ does not influence the value of $f$, and thus the solution of Eqs.(\ref{sistema simple 1-d} a-b) are those of the superconducting problem. We assume that they are non-zero and we will denote them as $(A_{\text{sc}},f_{\text{sc}})$.
	Furthermore, since $f_{\text{sc}}$ is non-zero, the superconducting term dominates the energy and turns off the nematic phase $g\sim 0$. This indicates that $\gamma_G$ is the parameter that must be adjusted to trigger the transition of $g$, from $g= 0$ to $g\sim 0$. Since this occurs for $\beta \to 0$, the critical value $\gamma_G^*$ will be obtained by increasing $\gamma_G\to +\infty$ until we find the transition. Analogously, $\gamma_F$ is the parameter that triggers the transition from $f=0$ to $f\sim 0$, and the critical value $\gamma_F^*$ is obtained by decreasing $\gamma_F \to 0$ until the transition is found. \bigskip 
	
	The test works as follows: 
	For $v>0$ fixed,
	(a) obtain the critical values for the \textit{zero-coupling} regime, $\gamma_F^*$ and $\gamma_G^*$, which means to obtain these critical values for the uncoupled problem; (b) solve the System \eqref{sistema simple 1-d} for different values of $\beta$ (using Procedure \ref{alg:main_solver}), and determine the values $\beta_1$ and $\beta_2$ at which the transitions $g\sim 0$ (for $\beta \to 0$) and $f\sim 0$ (for $\beta \to +\infty$) occur, respectively; (c) compare $\beta_1$ and $\beta_2$ with the reference values $\gamma_F^*/v$ and $v/\gamma_G^*$: since equality establishes the \textit{zero-coupling} condition, the weak coupling condition holds as long as $\gamma_F^*/v\sim \beta_1$ and $ v/\gamma_G^*\sim \beta_2$.
	
	It remains to indicate how both thresholds, $\gamma_F^*$ and $\gamma_G^*$, are obtained. 
	Starting with $g\sim 0$, we write $g=\delta \cdot G$ with $\delta \sim 0$ and replace it into Eq.(\ref{sistema simple 1-d}c). Taking $\delta\ll 1$, we get the asymptotic expressions $\delta^2 \left(\frac{4\pi^2\xi_n^2}{ \widetilde{d}^2}\right) G^2-1\sim -1$ and $f \sim f_{\text{sc}}$, where $f_{\text{sc}}$ is the superconductor response to the classical Ginzburg-Landau equation. The equation for $G$ is thus given by: 
	\begin{eqnarray}
		\label{eq:coefG}
		G^{\prime\prime}&=&\kappa_n ^2 G\left(4\pi^2\xi_n^2 \widetilde{d}^{-2} -1 +\gamma_G f_{\text{sc}}^2\right),\\ \nonumber
		G(0)&=&G(\widetilde{d})=0.
	\end{eqnarray}  
	The critical value $\gamma_G^*$ will be the \textit{largest} value of $\gamma_G$ such that Eq.\eqref{eq:coefG} has a non-zero solution.
	
	An identical situation arises for $\gamma_F^*$. We write $f=\delta \cdot F$, substitute it into Eq.(\ref{sistema simple 1-d}b), and take $\delta\ll 1$ to produce the asymptotic expressions $\delta^2 F^2-1\sim -1$ and $g \sim g_{\text{an}}$, where $g_{\text{an}}$ denotes the notrivial solution for the anharmonic oscillator. Then, the equation for $F$ is given by:
	\begin{eqnarray}\label{eq:gama_F}
		F^{\prime\prime}&=&\kappa_s ^2 F\left(B_e^2-1 +\gamma_F g_{\text{an}}^2\right)\; \\ \nonumber
		F^{\prime}(0)&=&F^{\prime}(\widetilde{d})=0.
	\end{eqnarray}
	
	The critical value $\gamma_F^*$ will be the \textit{smallest} value of $\gamma_F$ such that Eq.\eqref{eq:gama_F} has a non-zero solution.
	The values obtained for $B_e=1$ and $v=0.5$, rounded to six significant digits, are $\gamma_F^*=0.800056$ (corresponding to $f\sim 0$) and $\gamma_G^*=58.6509$. (corresponding to $g\sim 0$). The results for the consistency analysis are
	\[
	\begin{tabular}
		[c]{|l|c|c|}\hline
		& $g\sim0$ & $f\sim0$\\ \hline
		(A): uncoupl. probl. & $v/\gamma_{G}^{*}=0.008525$ & $\gamma_{F}^{*}/v=1.6001$\\\hline
		(B): actual problem & $\beta\sim0.009$ & $\beta\sim1.76$\\\hline
		(C): comparison & $5.57\times10^{-2}$ & $9.99\times10^{-2}$\\\hline
	\end{tabular}
	\]
	
	\noindent 
	where line (C) shows the relative error between lines (A), corresponding to the \textit{zero-coupling} regime, and (B), corresponding to the actual problem. Therefore, the transition values provided by the model, line (B), are close enough to the theoretical values computed with the zero-coupling, line (A), which we propose as a test to decide whether the weak coupling condition holds. With these results, we assure that the set of parameters involved in the numerical analysis performed in this work are consistent with a model that satisfies the weak coupling condition.
	
	\section{Conclusions}\label{V}
	
	In this article, we have investigated the interplay between superconductivity and nematicity in the context of the Ginzburg Landau theory. The Helmholtz Free Energy derived tries to uncover the behavior of the nematic and superconducting degrees of freedom in the presence of an external magnetic field, in notable difference from the case of an external electric field \cite{GarciaOvalle2020}. We apply the formalism to two different instances both bounded in one direction: an infinite cylinder with a circular cross-section and a strip; the latter being analyzed in detail by means of numerical calculations. After obtaining the corresponding Ginzburg-Landau equations, we have shown the existence of a threshold condition of a geometrical nature that must be satisfied to have a mixed response.\bigskip 
	
	Next, we have performed a numerical study of a slab under a special choice of parameters $v$, $\beta$, and $B_e$, in order to obtain nontrivial solutions. Our calculations suggest that for $B_{e}=1$ there are regions in the $(v,\beta)$ plane such that \textit{mixed solutions} exist. Actually, our computations suggest that for fixed $v$, the \textit{mixed solutions} exist only for $\beta$ within an open interval, denoted by $I(v)$, in which the left boundary corresponds to the transition $g\sim 0$ and the right one corresponds to the transition $f\sim 0$. In fact, for $v=0.5$, we have found an interval $I(0.5)\subset [0.008, 1.76]$ such that for $\beta \in I(0.5)$, the solution is mixed. In this case, we have noted that there is a transition from $f(x)\sim f_\text{sc}(x)$ (the solution for the superconductor case) and $g(x)\sim 0$ when $\beta \sim 0.009^+$, to $f(x)\sim 0$ and $g(x)\sim g_\text{an}(x)$ (the anharmonic-like solution) when $\beta \sim 1.75^-$. Furthermore, in all these cases, $f(x)$ and $g(x)$ displayed competition behavior, according to the positive sign of $v$, in which one of them concentrates while the other dilutes.\bigskip 
	
	Investigating the case $v<0$, we confirm a collaboration between nematicity and superconductivity and we also corroborate,  for fixed $v$, the existence of an interval $I(v)$ such that the \textit{mixed solutions} exist only for $\beta \in I(v)$. Working with $v=-0.02$, we obtain $I(-0.02)\subset [0.1,1.4]$. Remarkably, we note that the size of the interval is similar to the one for $v=0.5$, despite the fact that $v=-0.02$ is 20 times smaller. Besides, increasing the value of $|v|$ shrinks the length of the interval so as to get no \textit{mixed solution} at all for $v<-0.05$. Moreover, the critical values of $\beta$ are not related to the vanishing of $f(x)$ or $g(x)$, but to the presence of \textit{supersaturation} (which means that $1<\max |f|$ or $1<\max |g|$), indicating that the weak coupling condition is not satisfied.
	\bigskip 
	
	Regarding the dependence of the solutions on the magnetic field, we have computed the critical values for $\kappa=5$ and $d\xi_n^{-1}=20$, obtaining: $H_{c1}=0$ and $H_{c2}=4.99999$. Then, we have explored both $B_e>1$ and $B_{e}\sim 0$.
	Concerning the case $B_e>1$, we have shown that the critical value of the external magnetic field needed to cancel the superconducting parameter depends upon $(v,\beta)$, which we denote as  $H_{c2}^{*}(v,\beta)$. For $(v,\beta)=(0.5,0.8)$, we get the estimate $3<H_{c2}^{*}(0.5,0.8)<4$, meaning that for $B_{e}>H_{c2}^{*}(0.5,0.8)$, the superconductivity is destroyed and  $f\equiv 0$. Notice that $H_{c2}^{*}(0.5,0.8)/H_{c2}\sim 0.7$. We have also considered the case where $B_{e}\to 0^+$ with $(v,\beta)=(0.5,0.8)$, deducing that there exists a \textit{mixed solution} $f\not \equiv 1$ (which is a mixed superconductor-metallic state) and $g\not \equiv 0$. In addition, we have obtained a \textit{pure} superconducting solution $f\equiv 1$, working with $(v,\beta)=(0.5,0.5741)$ and $B_{e}=0$.\bigskip 
	
	Finally, we have considered the problem of deciding whether the chosen parameters $(v,\beta)$, with $v>0$, are consistent with a model that satisfies the weak coupling condition. To this end, we have designed a practical test, mainly based on energetic considerations, which consists of three steps: in first place, compute the critical values for the \textit{zero-coupling} model: $\gamma_F^*$ and $\gamma_G^*$, which are the values needed to get a nontrivial solution into the auxiliary linear model given by the uncoupled equations \eqref{eq:coefG} and \eqref{eq:gama_F}. Secondly, we solve the problem for $v>0$ using Procedure \ref{alg:main_solver}, and find the values for $\beta$ to get the transitions $g\sim 0$ and $f\sim 0$. Finally, we compare them to the ones provided by the \textit{zero-coupling} model previously obtained: $v/\gamma_G^{*}$ and $\gamma_F^*/v$. If these quantities are similar, we shall consider that the weak coupling condition is fulfilled. 
	
	We have applied this test to the set of parameters used throughout this article, verifying that for $0\leq v \leq 0.5$ and any $\beta \in [0.008,1.76]$ (with $B_e=1$, $\kappa=5$ and $d \xi^{-1}_n=20$), the corresponding model satisfy the weak coupling condition. For example, $v=0.5$ corresponds to $0.008\leq\nu|f_0|^{-1}\leq 1.76$, where $|f_0|=\frac{|a|^2}{2b}=\frac{H_c^2}{8\pi}$ is the magnitude of the Helmholtz free energy density of an infinite superconductor without nematicity and $H_c$ is the thermodynamic critical field \cite{Tinkham}, which can be directly measured in experiments.  
	


\end{document}